\documentclass[sigconf]{acmart}%
\usepackage[utf8]{inputenc}
\usepackage{tabularx}
\usepackage{color}
\newcommand\rev[1]{{{\color{black}#1}}}%
\author{Jonas Oppenlaender}
\email{jonas.oppenlaender@oulu.fi}
\orcid{0000-0002-2342-1540}
\affiliation{%
  \institution{University of Oulu}
  \city{Oulu}
  \country{Finland}
}%
\author{Elina Kuosmanen}
\orcid{0000-0001-6725-8848}
\email{elina.kuosmanen@oulu.fi}
\affiliation{%
  \institution{University of Oulu}
  \city{Oulu}
  \country{Finland}
}%
\author{Andr\'es Lucero}
\orcid{0000-0002-7176-2884}
\email{lucero@acm.org}
\affiliation{%
  \institution{Aalto University}
  \city{Espoo}
  \country{Finland}
}%
\author{Simo Hosio}
\orcid{0000-0002-9609-0965}
\email{simo.hosio@oulu.fi}
\affiliation{%
  \institution{University of Oulu}
  \city{Oulu}
  \country{Finland}
}%
\newcommand{\likertgraphswidth}{.758\linewidth}%
\newcommand{\asteriskheight}{.9cm}%
\newcolumntype{Y}{>{\centering\arraybackslash}X}
\newcommand\td[1]{\small{#1}}%
\copyrightyear{2021} 
\acmYear{2021} 
\setcopyright{acmlicensed}\acmConference[CHI '21]{CHI Conference on Human Factors in Computing Systems}{May 8--13, 2021}{Yokohama, Japan}
\acmBooktitle{CHI Conference on Human Factors in Computing Systems (CHI '21), May 8--13, 2021, Yokohama, Japan}
\acmPrice{15.00}
\acmDOI{10.1145/3411764.3445380}
\acmISBN{978-1-4503-8096-6/21/05}
\begin{document}%
%
\title[Comparing Crowdsourced and Peer Design Feedback]{%
    Hardhats and Bungaloos:
    Comparing Crowdsourced Design Feedback with Peer Design Feedback in the Classroom
}%
\begin{abstract}
Feedback is an important aspect of design education, and crowdsourcing has emerged as a convenient way to obtain feedback at scale.
In this paper, we investigate how crowdsourced design feedback compares to peer design feedback within a design-oriented HCI class and \rev{across two metrics: perceived quality and perceived fairness.} We also examine the perceived monetary value of crowdsourced feedback, which provides an interesting contrast to the typical requester-centric view of the value of labor on crowdsourcing platforms.
Our results reveal that the students ($N=106$) perceived the \rev{crowdsourced design feedback} as inferior to peer design feedback in multiple ways. However, they also identified various positive aspects of the online crowds that peers cannot provide. We discuss the meaning of the findings and provide suggestions for teachers in HCI and other researchers interested in crowd feedback systems on using crowds as a potential complement to peers.
\end{abstract}%
%
\renewcommand{\shortauthors}{J. Oppenlaender et al.}
%
\ccsdesc[500]{Information systems~Crowdsourcing}
\ccsdesc[500]{Human-centered computing~Empirical studies in HCI}
\ccsdesc[300]{Human-centered computing~Interface design prototyping}
\ccsdesc[500]{Human-centered computing~HCI design and evaluation methods}
%
\keywords{crowdsourcing, design feedback, crowd feedback system, classroom study, peer review}%
\begin{teaserfigure}%
 \centering%
 \includegraphics[height=5.6cm]{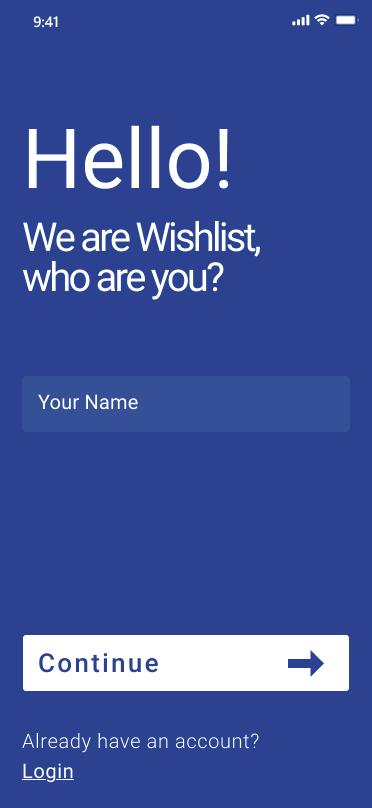}%
 \hspace{1mm}
 \includegraphics[height=5.6cm]{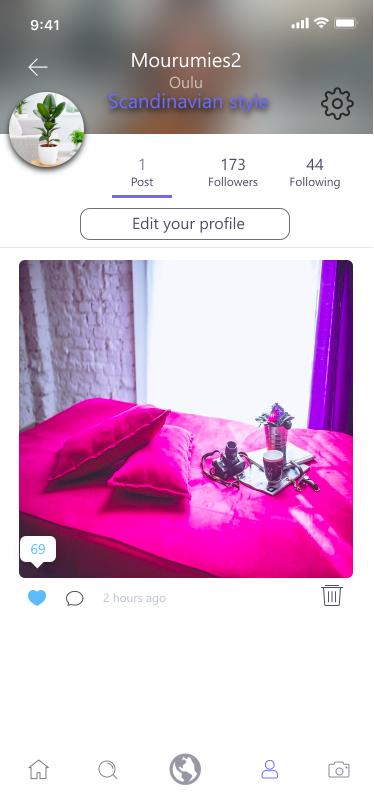}%
 \hspace{1mm}
 \includegraphics[height=5.6cm]{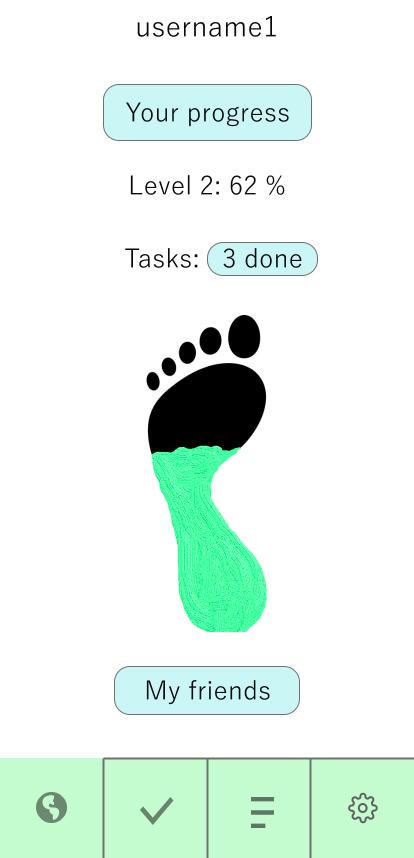}%
 \hspace{1mm}
 \includegraphics[height=5.6cm]{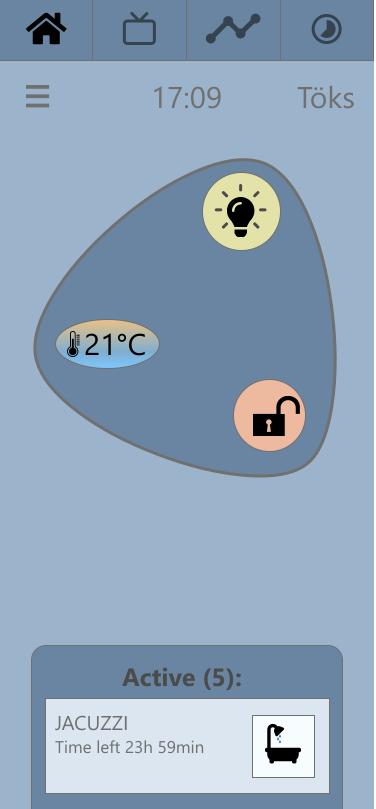}%
 \hspace{1mm}
 \includegraphics[height=5.6cm]{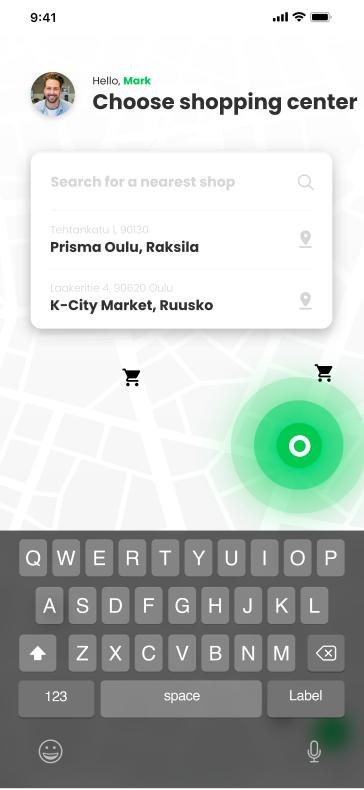}%
 \caption{Selected interactive prototypes of mobile applications created by students in the design course.}%
 \Description[Screenshots of interactive prototypes of mobile applications]{Five screenshots of selected interactive prototypes of mobile applications created in the design course.}%
\label{fig:screenshots}%
\end{teaserfigure}%
\maketitle%
%
%
\section{Introduction}%
\label{sec:introduction}%
%
Design is an important topic in HCI education.
While prior work has established the feasibility of using crowdsourced feedback in design education~\cite{crowdInnovationCourse-chi13.pdf,crowdInnovationCourse-hcomp2015.pdf,p5580-wauck.pdf},
\rev{not many
    deep investigations into the qualitative expectations and perceptions
    of students have been documented in the literature.}

\rev{Our work sets out to replicate and extend prior findings~\cite{crowdInnovationCourse-chi13.pdf,crowdInnovationCourse-hcomp2015.pdf,p5580-wauck.pdf}, particularly concerning the perceptions of the students.}
We provide a detailed empirical investigation into how students of a design-oriented undergraduate HCI course perceived and experienced crowdsourced 
design feedback, and how this feedback compared to peer design feedback in the classroom. In our study, students ($N=106$) were tasked with brainstorming, designing, and prototyping a functional online interface for a mobile application. Students received formative feedback on the interactive application interface from peers and crowd workers of a paid online crowdsourcing platform (Amazon Mechanical Turk or \textit{MTurk}).

\rev{Our study extends prior studies' findings with a comparison of the two sources of feedback across an extensive range of evaluation dimensions related to the perceived quality and felt experience of feedback.
We found that the perceived quality of the feedback is shaped by the perceived effectiveness, perceived effort, and the modality of the feedback.
Perceived fairness is shaped by agreeableness, valence (i.e., the affective tone and ``harshness'' of the feedback), diligence, usefulness, and credibility of the feedback.}
Additionally, we explore how students formulate a monetary valuation of crowdsourced design feedback, which contrasts the typical way of thinking about the monetary value of crowdsourced contributions as rewards paid by the requester.
This, we argue, may have implications for crowdsourcing setups in which the teacher as the requester of feedback (and thus the party who is setting the price for the crowdsourced task) is not the receiver and beneficiary of the crowdsourced feedback.

The main contributions of our work are:
\begin{enumerate}
    \item An empirical case study on using the crowd from Amazon Mechanical Turk as a source of design feedback in a university-level HCI course.
    \item A detailed analysis of the perceived attributes of formative design feedback from two sources: peers and the crowd from Amazon Mechanical Turk.
    \item \rev{An exploration of how students formulate a monetary valuation of crowdsourced design feedback.}
    \item A discussion on the implications and how teachers might best leverage crowdsourced design feedback in their courses.
\end{enumerate}

Overall, the students in our study found the formative design feedback from peers to be of higher quality across all of the analyzed criteria, except for valence. However, the diversity of people from different backgrounds was mentioned as a particularly positive aspect of the crowdsourced feedback, and approximately 25\% of the participants preferred MTurk feedback over peer feedback.
Concerning the monetary value of crowdsourced feedback, participant responses differed drastically from the pay that crowd workers typically receive on MTurk. Our findings are informative to researchers working on crowd feedback systems as well as teachers of design-oriented HCI courses who wish to explore crowdsourced feedback as a way to conveniently expose students to feedback from outside the course itself.%
%
%
%
\section{Related Work}%
\label{sec:related-work}%
\rev{The field of HCI has always had an interest in learning and education, for example by establishing a new dedicated CHI subcommittee on ``Learning, Education, and Families'' and a Special Interest Group (SIG) at CHI '19~\cite{Pammer2020}.}
Our work touches on 
design feedback, feedback collection via crowdsourcing, and the application of crowdsourced feedback in the classroom.%
%
%
\subsection{Design Feedback}%
Feedback is an important component of design education.
Feedback communicates a notion of a standard to the learner, and learners typically strive to minimize the feedback-standard discrepancy~\cite{e81ca813ed757e1e76c0023865c7dbdc7308.pdf}.
The overall goal of design education is to raise the learner's conception of standard to that of the teacher~\cite{sadler1989.pdf}.
One important mechanism to realize this goal is design feedback.

Design feedback can be given at four different levels (product, process, self-regulation, and self~\cite{Feedback.pdf}), and the feedback can be formative or summative.
While summative feedback is geared towards grading results at the end of a course, formative feedback guides students in improving their work~\cite{sadler1989.pdf}.
According to the theory of formative assessment~\cite{sadler1989.pdf}, formative feedback is instrumental to developing expertise.
Formative peer feedback for in-progress work may, for instance, improve course outcomes~\cite{Peerstudio.pdf}.
Effective feedback provides specific information about a learner's current performance together with explanations and concrete examples \cite{10677974.pdf}.
Successful feedback is specific, critical, and actionable \cite{sadler1989.pdf,Posts_paper_3.pdf}.

The term \textit{critique} is sometimes used as synonym for design feedback in related literature (e.g., in~\cite{p1024-nguyen.pdf}).
And indeed, a critique is ``the communication of a reasoned opinion about an artifact or a design''~\cite{p157-fischer.pdf}.
\rev{Traditionally, however, a (studio) critique refers to a formal, co-located feedback setting in which knowledgeable peers or experts provide feedback to student learners~\cite{dannels2008.pdf}. 
Critiques are a common practice in design education as a form of assessment. In a critique, co-located students discuss and evaluate their sketches, collages, or designs.
Critiques are also practiced among design colleagues as a means to receive feedback.}

In this work, we concentrate on product-based formative feedback for a design artifact. We follow a process-agnostic conceptualization of feedback as information provided by a \textit{feedback provider} (a teacher, peer, or other kind of \textit{reviewer} of a work) to the \textit{feedback receiver} (in our case, a university-level student)~\cite{Feedback.pdf}.%
%
%
\subsection{Crowdsourcing Design Feedback}%
\rev{Peer feedback has become a common pedagogical tool especially in online teaching and when the instructors are not always available~\cite{liaqat2020collaborating,elizondo2019quality}. Crowdsourcing offers an even more scalable approach to feedback. Researchers have in the past looked into the trade-offs of these approaches.}
\citeauthor{p1024-nguyen.pdf} found that design feedback from anonymous sources was perceived more positively compared to feedback from peers or an authority~\cite{p1024-nguyen.pdf}.
Anonymous feedback providers may give feedback that contains more specific criticism and praise, and thus be rated as more useful by the feedback receiver~\cite{crowdInnovationCourse-hcomp2015.pdf}.
Anonymity may further help minimize power differences between the feedback provider and the feedback receiver~\cite{dannels2008.pdf}.
To this end, crowdsourcing platforms, such as Amazon Mechanical Turk (\textit{MTurk}), may provide a fruitful ground for eliciting design feedback from a large and diverse group of anonymous people.
On MTurk, requesters (the feedback receivers) publish short ``Human Intelligence Tasks'' (\textit{HITs}) for anonymous workers (the feedback providers) to complete in exchange for a small monetary reward~\cite{howe2006}.

The microtask crowdsourcing model is, however,~-- by its nature~-- troubled with a number of issues that may negatively affect the provision of feedback. 
    Survey satisficing may inflate reliability and validity of collected data~\cite{10.1177_0013164415627349.pdf} and could negatively impact the quantity and quality of the collected feedback.
    The 
    power differences between workers and requesters
    manifest in low payment to workers \cite{data-driven-analysis-of-workers-earnings-on-amazon-mechanical-turk.pdf}.
    Workers may further be subject to numerous biases, such as the observer effect.
    On the other hand, feedback from peers may be positively biased~\cite{rightDesign.pdf}, and students were shown to appreciate and prefer feedback from external and anonymous sources~\cite{crowdInnovationCourse-chi13.pdf,p1024-nguyen.pdf}.

Researchers created a number of web-based systems to explore and investigate the potential of MTurk for collecting feedback on graphic designs.
\textit{Voyant}~\cite{p1433-xu.pdf} is a system for eliciting perception-based feedback on graphic designs from a non-expert crowd. The system aims to capture the first impressions and how well a graphic design meets its stated goals and design guidelines.
\rev{\textit{CrowdUI} \cite{CrowdUI} is a system for eliciting and aggregating visual feedback on the design of a website. The system enables users to modify a website's user interface. The user-generated design suggestions are presented to the feedback requester in aggregated form.}
\textit{CrowdCrit}~\cite{luther-crowdcrit-cscw2015.pdf}
gathers formative feedback on graphic designs from crowd workers and clients. Designers found the feedback provided by the system to be helpful and appreciated the specificity and level of detail of the feedback. Feedback from crowd workers was however found to be ``more generic.''
In \textit{Paragon}~\cite{paper606.pdf}, feedback providers complement written feedback on graphic designs with visual design examples.
Feedback provided in this manner was found to be more novel, specific, and actionable.
\textit{SIMPLEX}~\cite{SIMPLEX} is a 
system to gather summative feedback for designs and artworks from a situated crowd using two public displays.
\textit{ZIPT}~\cite{p727-deka.pdf} is a system  that~-- unlike the above systems that elicit feedback on static graphic designs~-- uses virtualization technology to enable a crowd to conduct remote user tests on mobile applications. This setup is similar to the one used in our study, except that feedback providers in our case provided feedback on a high-fidelity web-based prototype, not a real mobile application.

Researchers further investigated strategies for improving the quality of crowdsourced design feedback.
Rubrics 
are an effective way to structure feedback and raise its perceived value, valence (i.e., affective tone), and specificity~\cite{Posts_paper_3.pdf}.
Scaffolds were shown to support students in reflecting on the feedback received~\cite{3357236.3395480.pdf}.
Structured workflows were found to generate feedback that was more  interpretative, diverse and critical than free-form prompts~\cite{p1637-xu.pdf}.
Further, guiding questions
formulated by the feedback receiver were found to be effective scaffolds to facilitate the exchange of feedback among peers~\cite{paper138.pdf}.
Measures such as the above have been found to improve the quality of crowdsourced design feedback.%

In contrast to the above quality measures, our study provides insights into the felt experience of evaluating feedback.
The participants in our study were given the responses from Amazon Mechanical Turk \textit{as is}, without any filtering of responses or other post-hoc measures for improving data quality.
We deliberately chose to compare the ``unfiltered'' feedback from the two sources. \rev{We decided against filtering the crowdsourced data as we believe this is the fairest way of comparing the two feedback sources without paying favors to one source. This decision avoids introducing bias to the study and best reflects what can be expected from the two types of feedback providers.} However, we did employ standard qualification measures widely used in academic studies on MTurk.
%
%
%
\subsection{Crowdsourced Feedback in the Classroom}%
%
\citeauthor{crowdInnovationCourse-chi13.pdf} established the feasibility of using crowds from MTurk for collecting design feedback in the classroom~\cite{crowdInnovationCourse-chi13.pdf}.
In \citeauthor{crowdInnovationCourse-chi13.pdf}'s studies, the crowd provided feedback along four key stages of the innovation process (needfinding, ideating, testing, and pitching).
Students found feedback from crowd workers to be beneficial in all four stages and the crowdsourced feedback helped students ground their design efforts in real-world opportunities. Most relevant to our study, testing early-stage storyboards with the online crowd was found to help students uncover issues and directions for future work.

\citeauthor{p1637-xu.pdf} used a crowd feedback system 
to provide 10~students in a visual design course with formative feedback from workers on MTurk~\cite{p1637-xu.pdf}.
The authors found that formative feedback from the crowd prompted students to change and improve their designs.
Compared with the feedback from experts, the crowdsourced feedback was however not found to agree on whether a design met its communicative goals.

\rev{In the work by \citeauthor{p5580-wauck.pdf}, students created prototypes of user interfaces in a project-based design course.
Students received feedback from their peers and three different online crowds: the students' own social network, online communities, and Amazon Mechanical Turk.
The authors measured the quality, quantity, and valence of the feedback, and how it was acted upon.
Student peers were found to provide feedback that was higher in perceived quality, more acted upon, and longer than feedback from online communities.
Further, summative feedback 
from both peers and online communities was found to be more negative in valence than formative feedback at an early stage in the design project.}

\rev{Our study is similar to the study by \citeauthor{p5580-wauck.pdf}, both in its focus and study design.
Our study is therefore partially a replication study of the work by \citeauthor{p5580-wauck.pdf}.
Our contribution to HCI is the confirmation of the prior study's findings~\cite{REPLICATION}.
The study by \citeauthor{p5580-wauck.pdf} answers three research questions, only one of which is student perceptions, whereas our work focuses only on the student perceptions, but extends the prior study to more dimensions qualitatively and quantitatively.
Our work therefore extends and contributes a clear increment to prior studies of feedback in the classroom with an in-depth analysis of students perceptions.}

Prior literature found that
student designers may attribute value to feedback along a number of different criteria, such as the feedback's
    %
    %
    \textit{quantity}~\cite{p5580-wauck.pdf,Posts_paper_3.pdf,luther-crowdcrit-cscw2015.pdf},
    \textit{specificity}~\cite{sadler1989.pdf,Posts_paper_3.pdf,crowdInnovationCourse-hcomp2015.pdf,luther-crowdcrit-cscw2015.pdf,paper606.pdf,3357236.3395480.pdf,paper138.pdf}, 
    \textit{criticality}~\cite{sadler1989.pdf,Posts_paper_3.pdf,3357236.3395480.pdf,paper138.pdf},
    \textit{valence}, \textit{affect}, or \textit{sentiment}~\cite{p5580-wauck.pdf,Posts_paper_3.pdf,p1024-nguyen.pdf,p1637-xu.pdf,paper138.pdf},
    \textit{helpfulness} \cite{p1433-xu.pdf,luther-crowdcrit-cscw2015.pdf,p727-deka.pdf,p4627-krause.pdf}, 
    %
    %
    %
    \textit{fairness}~\cite{Posts_paper_3.pdf}, and
    \textit{actionability}~\cite{sadler1989.pdf,p5580-wauck.pdf,Posts_paper_3.pdf,paper606.pdf,3357236.3395480.pdf,paper138.pdf}.
    %
    Students may, for instance, find specific and emotionally positive feedback useful and
    students are likely to appreciate longer and more actionable feedback with clear justifications~\cite{Posts_paper_3.pdf}.
In our study, we provide a mixed-method investigation into these aspects of formative design feedback in the context of a university-level design course.%
%
%
%
%
%
\section{The Study}%
\label{sec:Study}%
Our aim was to investigate how students perceive and experience crowdsourced formative feedback in the classroom and 
how this feedback compares to peer feedback from the students' classmates.
The study was conducted in the context of a 
design course
at the University of Oulu, Finland.

\subsection{HCI Course and Design Projects}%
\label{sec-sub:HCI-course}%
%
The study was conducted in a 9-week undergraduate course
(521145A ``Human Computer Interaction'')
between October and December 2019.
The course followed a project-based learning approach.
In the beginning of the course, students formed 42~groups with three members each around self-chosen topics.
Each group of students was tasked to brainstorm, design, and prototype a user interface of a mobile application.
Each group first sketched their idea with pen and paper and evaluated this paper prototype in a user study with their classmates. The groups then transferred their designs into a digital 
prototype using \textit{Adobe~XD} (a user interface design tool), while taking into account the lessons-learned from the user studies.
The interactive prototypes were published online using Adobe~XD's functionalities, which
    allowed us to invite both students and workers from a crowdsourcing platform to operate and review the interactive application prototypes.
Each group received formative feedback from other students and crowd workers.
After reviewing the formative feedback, the groups acted on the feedback received and submitted a revised version of the digital prototype for grading.
Selected examples of the interactive designs created during the course are depicted in Figure~\ref{fig:screenshots}.

\begin{figure*}
\centering%
\includegraphics[width=.721\linewidth]{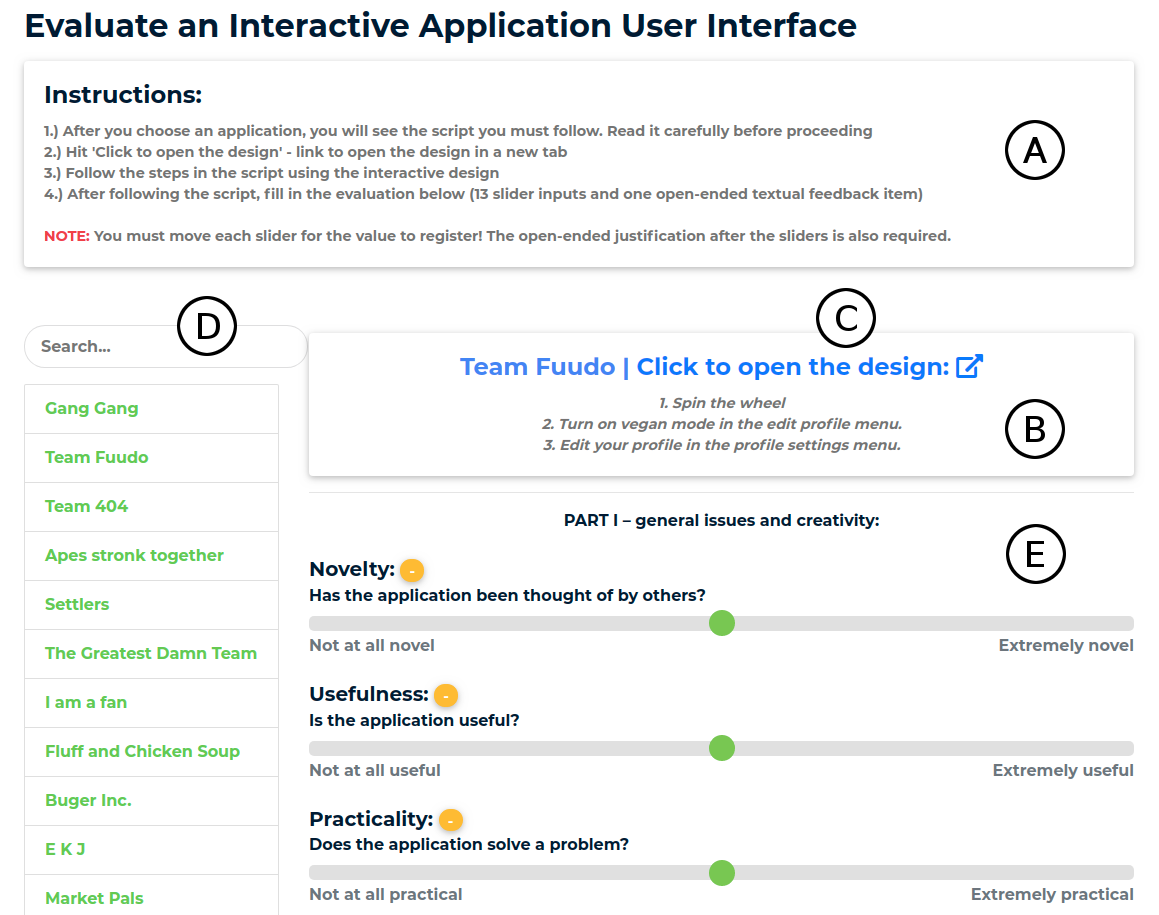}%
\caption{Web-based interface for feedback collection.
A set of general instructions was provided to the reviewers (A).
Each student group provided their own set of tasks to be completed in the mobile application (B).
A click on a link (C) directed the reviewer to the interactive prototype of the mobile application hosted on Adobe~XD resource servers.
The reviewer was instructed to rate the mobile application along a number of dimensions (E).
Students could access the groups they were assigned to review in the sidebar (D). 
A screenshot of the full interface is available in the auxiliary material.
}%
\Description{A screenshot depicting the web-based graphical user interface used for collecting feedback from students and Amazon Mechanical Turk workers. Feedback was collected in two forms: with sliders and open-ended text entry.}%
\label{fig:interface}%
\end{figure*}%

\subsection{Study Design}%
\label{sec-sub:study-design}%
\rev{Our study design choices were shaped by practical and pedagogical considerations and informed by our study being a live, authentic teaching exercise rather than a controlled laboratory experiment on feedback quality.
Our study therefore needs to be viewed in light of being a case study of feedback with ecological validity in the classroom.}

We conducted a within-subject \rev{study} in the context of the HCI course. Each of the 42~groups of students was provided a zip file (the ``feedback package'') with formative feedback from two sources: nine of their peers and nine crowd workers.
\rev{The number of feedback items was determined based on the group size of the students. Since each group consisted of three members, we asked each student to review the work of three other teams.}
Students were briefly introduced to the notion of crowdsourcing in the course lectures.
The within-subject design was 
mandated by the university's teaching rules, to provide each group of students a similar learning experience and not to disadvantage some groups in the graded course.
Students were individually asked to comment on the formative feedback in a
survey (the ``final questionnaire'') at the end of the course.
In the following section, we describe the procedure for collecting feedback, how feedback was provided to the students, and the final questionnaire.

\subsubsection{Apparatus and materials}%
\label{sec-sub:apparatus-materials}%


Feedback was collected using the \rev{web-based survey instrument by
\citeauthor{Hosio:2016:LWC:3056355.3056393}~\cite{Hosio:2016:LWC:3056355.3056393}} (depicted in Figure~\ref{fig:interface}).
The interface included 14~questions, of which 13~were elicited with sliders.
Both students and crowd workers (henceforth: the \textit{reviewers}) received the same instructions and provided feedback with the same interface, with one minor difference.
Students were presented with a list of all groups and navigated to the group they were assigned to review. For the review by crowd workers, the sidebar with group names was hidden to not distract the workers. Instead, the task was set up so that the worker would only see the one design that was to be rated during that task. 
For the students, providing peer feedback was a fixed part of the coursework, and students provided feedback individually in their self-study hours outside the scheduled exercises.


Using the web-based interface, the reviewers rated the mobile application prototype on several criteria (each on a 7-point anchored Likert scale; see Figure~\ref{fig:interface}).
We elicited the {novelty} and {usefulness} of the application prototype as two important components of product-based
creativity~\cite{2012RuncoJaegerStandardDefinition.pdf}. 
We further elicited {practicality} and a rating of the {design}.
Each criteria was explained with a short question. For instance, novelty was explained with ``Has the application been thought of by others?''
Reviewers were asked to rate their {success} in following the set of tasks provided by the student group.
Further, the interface included the 8-item version of the User Experience Questionnaire (UEQ-S)~\cite{UEQ-S}.
The UEQ-S measures several aspects of user experience on dichotomous scales, such as ``Boring-Exciting,'' ``Confusing-Clear,'' and ``Complicated-Easy.''
In the last item, reviewers were asked to provide an exhaustive open-ended justification for how they scored the different criteria.


{
\begin{table*}
\caption{Feedback package with peer and crowdsourced feedback provided to the students.}%
\label{tab:zipcontents}%
\begin{tabularx}{\textwidth}{lXXXX}%
\toprule%
    \td{\textbf{File}} &
    \td{\texttt{data.xlsx}} & 
    \td{\texttt{summary.png}} &
    \td{\texttt{summary\_openended\_ class.txt}} &
    \td{\texttt{summary\_openended\_ mturk.txt}} \\
\midrule
        \td{\textbf{Description}}
        & 
        \td{Quantitative responses (raw data) collected with the survey instrument}
        &
        \td{Graphical summary (bar chart) of the feedback, split between peer and MTurk feedback}
        &
        \td{Open-ended feedback from student peers}
        &
        \td{Open-ended feedback from workers on MTurk}
        \\
\bottomrule
\end{tabularx}
\end{table*}
}

\paragraph{Feedback Package}
The collected feedback from the reviewers was bundled in a zip file (see Table~\ref{tab:zipcontents}).
Each feedback package contained raw data collected with the web-based survey instrument, a graphical summary of Likert-scale responses (split between peer and crowd feedback), and two text files with the open-ended feedback from nine student reviewers and nine crowd workers, respectively.
The feedback package was accompanied with detailed instructions for the students. 
A full example of a feedback package is available in the auxiliary material.


\subsubsection{Study procedure}%
\label{sec-sub:procedure}%
\rev{Students individually inspected the feedback package at the end of the course (in their self-study hours outside of the scheduled exercises). Feedback was not given in a specific order, and the students could freely choose which one to consume first.
The students then evaluated the contents of the feedback package in an online questionnaire.}
All related ethical procedures were followed as required by our University.
Students were asked to consent to their data being used for the purpose of the academic study prior to completing the study. Students were informed that declining consent did not affect their course grades.
Students were specifically instructed to fill out the final questionnaire individually, not in teams.
As an incentive, 5~points counting towards the course credit were awarded to the students who calculated the three components of the UEQ-S (Overall, Pragmatic, and Hedonic Quality) from the raw data.

The final questionnaire consisted of 18~items including demographic questions, an evaluation of the formative feedback, and the satisfaction with the course.
The questionnaire items related to the evaluation of the feedback are listed in tables~\ref{tab:likertquestions} and \ref{tab:codedquestions}.
The full final questionnaire can be found in the auxiliary material.

The questionnaire elicited detailed justifications for the students' preference of feedback (see Table~\ref{tab:codedquestions}).
We first inquired which of the two feedback sources the students preferred overall.
This first open-ended item was given to capture students' reasoning in their own terms, without imposing a structure or limiting the response.
The following three items were structured around three principles of good feedback: \textit{effectiveness}, \textit{fairness}, \textit{actionability}~\cite{sadler1989.pdf,Posts_paper_3.pdf}.
Students were next asked to quantitatively rate the feedback on 7-point Likert scales (see Table~\ref{tab:likertquestions}). Effectiveness was measured with items related to effective feedback~\cite{10677974.pdf}: \textit{specificity}, \textit{actionability}, and \textit{explanations}. \textit{Fairness} and \textit{actionability} had their own items, and in addition, students were asked to judge the overall \textit{valence} of each feedback source, the \textit{relevance} of the feedback to the application prototype, the overall \textit{quality} of the feedback, and the \textit{satisfaction} with the respective feedback source.
Finally, students were asked to provide their personal estimate of the \textit{monetary value} of the feedback provided by crowd workers (\textit{``Looking only at the feedback from online workers on Mechanical Turk for your own group, how much do you think this feedback is worth, in money (Euros)?''}).
Students were specifically instructed to estimate the cost of \textit{``all of the feedback from MTurk combined (but for your own group)~-- not for an individual feedback item among the many.''}

\subsection{Participants}%
\label{sec-sub:participants}%

\subsubsection{Crowd workers}%
\label{sec-subsub:crowd-workers}%
We invited crowd workers from Amazon Mechanical Turk to review the 42~mobile application prototypes created by the student groups. Each prototype was assigned to nine different crowd workers.
Workers were recruited following best practices (HIT approval rate greater than 95\% and number of HITs approved greater than 100). Similar qualification criteria are widely used in academic studies~\cite{peer2013.pdf}.
In total, 387~HITs (with an attrition rate of 2.3\%) were completed by 328~unique workers.
Workers were paid US\$ 1 per HIT completed.

\subsubsection{Students}%
\label{sec-subsub:students}%
Out of 112~students who submitted the final questionnaire,
106~students (P1--P106, aged from 18~to 58, $M=23.3$~years, 
$SD=5.6$~years) consented to contribute their data to the study.
As the course was offered by 
    a Faculty of Technology,
the gender distribution of the participants reflects the bias that is often found in technology-dominated subject areas (82~male, 20~female, 1~non-binary/third gender, and 3~undisclosed).
The course was primarily aimed at 
Bachelor students (78~participants), but Master students were allowed to attend (28~participants).
Over half the students (62 participants) were enrolled in a degree program in Computer Science. However, the course
was also attended by students from other disciplines, such as 
  Industrial Engineering (24 participants),
  Electrical Engineering (12 participants),
  English Philology (1 participant),
  Medicine (1 participant), and 
  other degree programs (6 participants).


\begin{table*}[htb]%
\centering%
\caption{Likert-scale items in the final questionnaire given to students at the end of the course. The Likert plots reflect the percentage of participants who agreed, disagreed, or were neutral to a given statement. For example, 62\% of the students thought the formative feedback from MTurk workers was not specific, 7\% neither agreed nor disagreed, and 31\% agreed to this statement.
}%
\label{tab:likertquestions}%
\begin{tabularx}{\textwidth}{%
        >{\hsize=2\hsize}X  |
        >{\hsize=.3\hsize}Y |
        >{\hsize=.3\hsize}Y |
        l
    }
    \toprule
    \td{\bf Feedback item}
        & \td{\bf Peer feedback\newline M~(SD)}
        & \td{\bf Crowd feedback\newline M~(SD)}
        & \td{\bf Wilcoxon rank sum} \\
    \midrule
        \td{The feedback is specific\textsuperscript{1}}
        & \td{5.08 (1.28)} & \td{3.31 (1.72)}
        & \td{
            $W=3897.5$  
        } \\
        \multicolumn{4}{l}{
            \includegraphics[width=\likertgraphswidth]{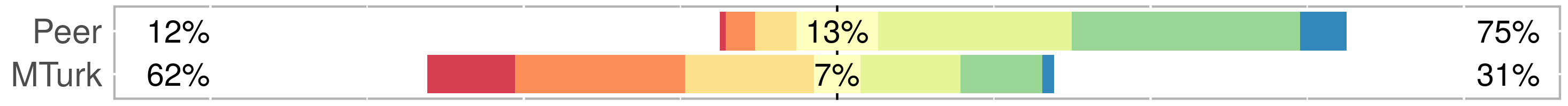}%
            {%
                \includegraphics[height=\asteriskheight]{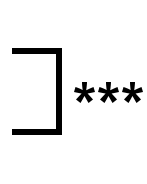}%
            }%
        } 
        \\
\midrule

        \td{The feedback is actionable\textsuperscript{1}} & \td{4.95 (1.36)} & \td{3.34 (1.78)}
        & \td{
            $W=3482$    
        } \\
        \multicolumn{4}{l}{
            \includegraphics[width=\likertgraphswidth]{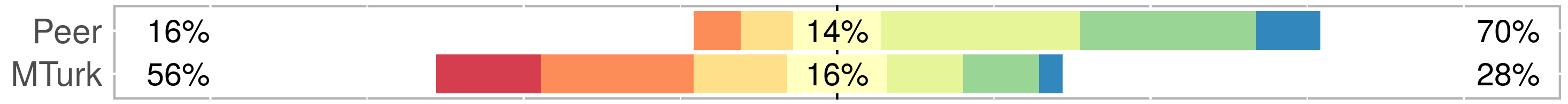}%
            {%
                \includegraphics[height=\asteriskheight]{figures/asterisk_3.png}%
            }%
        }
        \\
\midrule

        \td{The feedback contains explanation(s)\textsuperscript{1}} & \td{4.65 (1.38)} & \td{3.21 (1.80)}
        & \td{
            $W=3542$    
        } \\
        \multicolumn{4}{l}{
            \includegraphics[width=\likertgraphswidth]{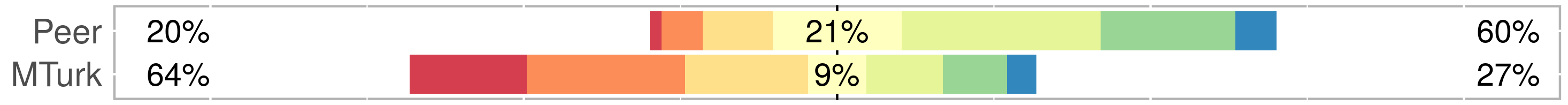}%
            {%
                \includegraphics[height=\asteriskheight]{figures/asterisk_3.png}%
            }%
        }
        \\
\midrule

        \td{Please rate the overall valence (positivity/negativity) of the feedback from the two feedback sources.\textsuperscript{2}} & \td{4.49 (1.29)} & \td{4.51 (1.62)}
        & \td{
            $W=2036.5$  
        } \\
        \multicolumn{4}{l}{
            \includegraphics[width=\likertgraphswidth]{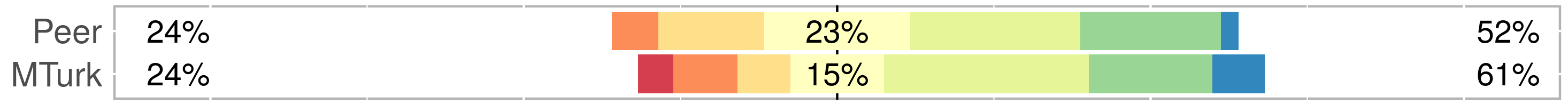}%
        }
        \\
\midrule

        \td{Please rate the relevance of the two feedback sources to your application prototype.\textsuperscript{3}} & \td{4.91 (1.16)} & \td{3.53 (1.58)}
        & \td{
            $W=3505.5$  
        } \\
        \multicolumn{4}{l}{
            \includegraphics[width=\likertgraphswidth]{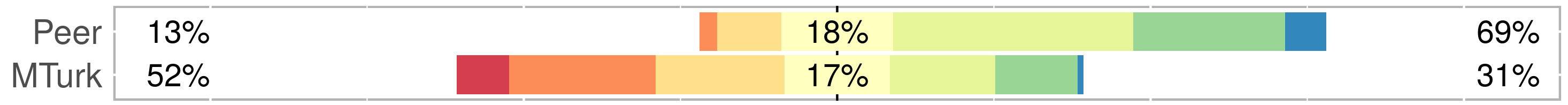}%
            {%
                \includegraphics[height=\asteriskheight]{figures/asterisk_3.png}%
            }%
        }
        \\
\midrule

        \td{Please rate the overall quality of the two feedback sources:\textsuperscript{3}} & \td{4.74 (1.20)} & \td{3.35 (1.70)}
        & \td{
            $W=8435.5$  
        } \\
        \multicolumn{4}{l}{
            \includegraphics[width=\likertgraphswidth]{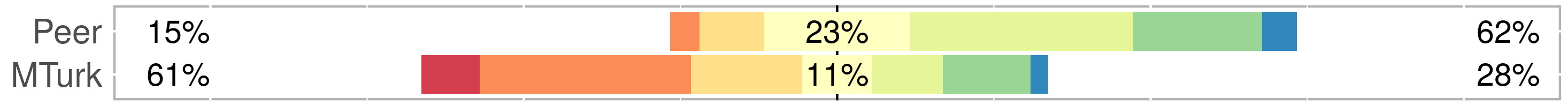}%
            {%
                \includegraphics[height=\asteriskheight]{figures/asterisk_3.png}%
            }%
        }
        \\
\midrule

        \td{Please rate your own satisfaction with the two feedback sources.\textsuperscript{4}} & \td{5.32 (1.51)} & \td{3.55 (2.06)}
        & \td{
            $W=8480$    
        } \\
        \multicolumn{4}{l}{
            \includegraphics[width=\likertgraphswidth]{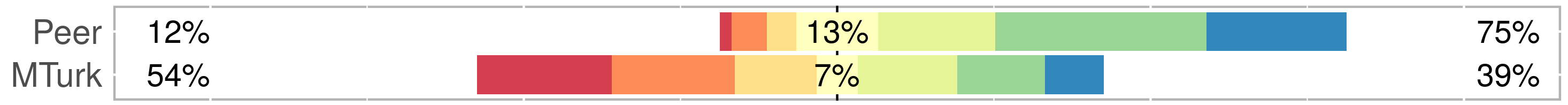}%
            {%
                \includegraphics[height=\asteriskheight]{figures/asterisk_3.png}%
            }%
        }
        \\



    \bottomrule
  \end{tabularx}
  \\[0.5\baselineskip]
  \begin{minipage}{1\textwidth}
   \footnotesize
   \includegraphics[width=5cm]{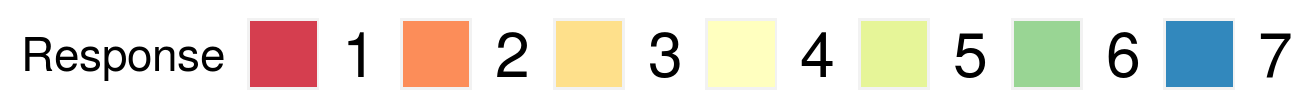} \\
   \textsuperscript{1} 1~-- Strongly Disagree ... 7~-- Strongly Agree \\
   \textsuperscript{2} 1~-- Strongly Negative ... 7~-- Strongly Positive \\
   \textsuperscript{3} {1~-- Very poor,
            2~-- Poor,
            3~-- Fair,
            4~-- Good,
            5~-- Very good,
            6~-- Excellent,
            7~-- Exceptional} \\
   \textsuperscript{4} {1~-- Very dissatisfied,
            2~-- Moderately dissatisfied,
            3~-- Slightly dissatisfied,
            4~-- Neutral,
            5~-- Slightly satisfied,
            6~-- Moderately satisfied,
            7~-- Very satisfied} \\
   \textsuperscript{***} p < 0.001
  \end{minipage}
\end{table*}

\subsection{Methodology}%
\label{sec-sub:methodology}%

We focus our mixed-method analysis on the students' evaluation of the feedback in the final questionnaire. We only briefly report on the feedback itself, because we want to foreground the subjective experience of formative design feedback in this paper. 

We compared different criteria of how students perceived the feedback, as described in Section~\ref{sec-sub:procedure}.
Since the data was not following a normal distribution (according to Shapiro-Wilk's test), we used paired Wilcoxon rank sum tests to evaluate the differences in the students' perception of the peer feedback and the crowdsourced feedback.
Besides investigating the students' perspective on the feedback, we quantitatively analyzed and compared the open-ended feedback along three criteria contributing to the quality of the feedback:
    the \textit{length} of the feedback (measured by the number of characters in the feedback item),
    the amount of 
    noise in the feedback (as measured by the presence of unrelated and nonsensical words), and
    the \textit{effort} spent on providing the feedback (as measured by the time taken to complete both the Likert-scale feedback and the open-ended feedback). 


Qualitatively, 
the three open-ended items (see Table~\ref{tab:codedquestions}) were analyzed following the guidelines for content analysis~\cite{weber1990basic}.
One researcher first familiarized himself with the data and extracted verbatim terms from the first item of the questionnaire in which students provided reasons for their overall preference of feedback.
The researcher consolidated the extracted terms into codes, while also considering the responses to the other two open-ended questionnaire items about the effectiveness and fairness of the feedback (Table~\ref{tab:codedquestions}).
Each code was categorized on whether it was a positive or negative statement about peer or crowd feedback, respectively.
Subsequently, the researcher iteratively and inductively grouped the codes into themes and sub-themes.
Next, a codebook was developed with descriptions for each code. After discussing the codebook, the first and second authors of this paper and one additional student individually coded all responses to the first open-ended item of the questionnaire.
Inter-rater reliability improved after reconciling differences in two discussions and adjusting the codebook. The final inter-rater reliability among the three raters (as measured by Fleiss Kappa~\cite{395-1971_Fleiss0001.pdf}) was \rev{$\kappa = 0.83$.} 

\section{Quantitative Comparison of Peer and Crowdsourced Feedback}


We found significant differences in how the students perceived the feedback from peers and crowd workers in all criteria except valence (see Table~\ref{tab:likertquestions} on page~\pageref{tab:likertquestions}).

Peer feedback was significantly more \textit{specific}, more \textit{actionable}, and contained more \textit{explanations} than the crowdsourced feedback in the opinion of the students (each $p<0.001$).
Peer feedback was also deemed more \textit{relevant} to the mobile application prototype and of better \textit{quality} than the crowdsourced feedback (each $p<0.001$).
Overall, students were more \textit{satisfied} with the peer feedback compared to the crowdsourced feedback from MTurk ($p<0.001$).
On average, the feedback from MTurk was perceived slightly more positive in affective tone than the peer feedback ($M=4.51$ for crowdsourced feedback versus $M=4.49$ for peer feedback). However, no statistical difference between peer feedback and crowdsourced feedback was found in terms of \textit{valence} ($p>0.05$).


An inspection of the feedback revealed that some of the answers to the open-ended item were extremely short and contained either non-existing words (e.g., \textit{``uehi,''} \textit{``asdf,''} and \textit{``adse''}), only numbers, or other words not relevant to the task (such as \textit{``yes,''} \textit{``no,''} \textit{``this app,''} 
and \textit{``NICE''}).
We manually curated a list of words and found the amount of 
noise greatly differed between the two feedback sources.
Only 0.3\% ($N=1$) of the open-ended peer feedback contained 
noise, compared to 12.2\% ($N=47$) of the feedback from MTurk.

Related to the amount of 
noise, the peer feedback contained significantly longer explanations compared to the crowdsourced feedback (peer feedback $M=296$ characters, $SD=287$ characters versus crowdsourced feedback $M=117$ characters, $SD=173$ characters, $p<0.001$, $W = 105454.5$).
The amount of answers with less than 10~characters (including the noise) 
was 0.6\% in the peer feedback and 23.4\% in the crowdsourced feedback.


\newcounter{itemnum}%
\begin{table*}
\caption{Qualitative items of the final questionnaire given to students at the end of the course, with coded preference for peer feedback and crowd feedback. For example, 79~students overall preferred the peer feedback, 26~the crowdsourced feedback, and one student could not decide.}%
\label{tab:codedquestions}%
\begin{tabularx}{\textwidth}{Xlllll}%
\toprule
     \td{\textbf{Item}} &
     \td{\textbf{Type}} & \td{\textbf{Peer}} & \td{\textbf{MTurk}} & \td{\textbf{undecided}} & \td{\textbf{n/a}\textsuperscript{1}} \\
\midrule
    \td{Which of the two feedback sources (in-class from other students or crowdsourcing from online workers) do you \textbf{prefer}, and why?} &
    \td{Open-ended} & \td{79} & \td{26} & \td{1} & \td{--} \\
    \td{How \textbf{effective} did you find the feedback (from both sources combined) in helping you understand what you did well and how you could do better?} & 
    \td{Open-ended} & \td{--} & \td{--} & \td{--} & \td{--} \\
    \td{Which of the two feedback sources (in-class or crowdsourcing)  do you think was more \textbf{fair},
        and why?} & 
    \td{Open-ended} & \td{62} & \td{34} & \td{9} & \td{1} \\
\midrule
    \td{Which of the two feedback sources (in-class or crowdsourcing) contained items that you would be able to act on -- that is, which feedback is more \textbf{actionable}?} &
    \td{Multiple-choice\textsuperscript{2}} & \td{84} & \td{22} & \td{--} & \td{--} \\
\bottomrule
\end{tabularx}
    \\[0.5\baselineskip]
    \begin{minipage}{0.95\textwidth}
        \footnotesize
        \textsuperscript{1} Preference could not be determined from the response. \\
        \textsuperscript{2} ``In-class feedback from other students'' or ``MTurk feedback from online workers'' \\
    \end{minipage}
\end{table*}

The time spent on writing the feedback varied between 18.2~seconds and 22~hours.
As the extremely long times do not reflect the actual time spent on the task, we removed times longer than one hour in the following comparison.
\rev{One hour was the 
the maximum time available for a crowd worker to complete the survey in our data collection campaign on MTurk.
There were 22~observations over 1~hour in the peer feedback,
    and none in the crowdsourced feedback.}
On average, the class took more time to provide feedback ($M=12.2$~min, $Min=18.3$~seconds, $Max=54.2$ min, $SD=10.2$ min) than the crowd workers ($M=4.6$~min, $Min=18.2$~seconds, $Max=42.2$~min, $SD=5.2$~min; $p<0.001$, $W=103461.5$).

\section{Qualitative Findings}%
%
We coded the open-ended questions to determine the preference of each student for either peer or crowdsourced feedback (see Table~\ref{tab:codedquestions}).
The majority of the students ($N=79$; 74.5\%) voiced a preference for peer feedback, but a not entirely insignificant minority ($N=26$; 24.5\%) preferred the feedback from MTurk.
Overall, students found the peer feedback more \textit{useful}, more \textit{effective}, more \textit{actionable}, and of greater \textit{fairness}. In contrast, students often noticed and complained about the low perceived \textit{effort} in the crowdsourced feedback.
We elaborate on each of these aspects in the following sections.
We structure our qualitative findings around three main 
themes of experiencing feedback:
    \textit{feedback quality}
        (with sub-themes
            \textit{effectiveness},
            \textit{effort}, and
            \textit{modality}),
    \textit{feedback fairness}
        (with sub-themes
            \textit{agreeableness},
            \textit{valence},
            \textit{diligence},
            \textit{usefulness},
            and \textit{credibility}),
and the \textit{monetary valuation} of the feedback.

\subsection{Perceived Quality of the Feedback}%

\subsubsection{Perceived effectiveness of the feedback}%

The overall perception of the effectiveness of the feedback (from both sources) ranged from \textit{``not that effective''} (P3) to \textit{``highly effective''} (P5).

Peer feedback was overall perceived as more meaningful and comprehensive. Peers provided more constructive and actionable suggestions for improvement than the crowdsourced feedback.
Several students lauded the peer feedback for \textit{``pinpointing design flaws''} (P18) and specific elements to improve, such as buttons, font sizes, wording, and other \textit{``little things that bugged [the reviewers]''} (P5).



The open-ended crowdsourced feedback, on the other hand, was seen as being more \textit{``general''} (e.g., P7) and, as a consequence, less useful and actionable by the majority of students.
Several students complained about the low number of takeaways found in the crowdsourced feedback.

Only few students ($N=9$) found both sources equally effective. P8, for instance, thought that \textit{``the class noticed the bad sides and the online workers noticed the good sides of the app.''}
P21 searched the fault in his own group, as he thought that
    \textit{``the design was not realistic and attractive for a general public, and some features were not that clear''}, but expressed optimism that \textit{``after [implementing the review] we think that the opinions would change.''} 

\subsubsection{Perceived effort}%

Students noticed a difference in effort put into writing the open-ended feedback.
The peer feedback was perceived as being more elaborate.
The length of the feedback affected the perception of usefulness, as students were able to find fewer takeaways in the crowdsourced feedback.
Crowdsourced feedback was often perceived as nonsensical, primarily due to some of the responses being extremely short and unrelated to the task:
    \textit{``My group didn't learn anything useful from online workers open ended questions, it had answers like "24", "god like app" and "Useful app :)". All of our classmates answered with many rows of text''} (P16).

Students recognized that their peers had put more thought and time into writing the feedback, whereas the
    \textit{``online workers just tried to speedrun the questions''} (P91).
Two students suspected the feedback by MTurk workers was provided by \textit{``bots and other ESLs who probably didn't even give it anything but a glance to get their pennies''} (P3), and that \textit{``the slider selections were done randomly by these bots''} (P1).

\subsubsection{Modality of the feedback}%


Students generally preferred the open-ended feedback over the raw data and summary chart.
The written justifications allowed the students to identify specific problems in their user interfaces and areas to improve.
%
%
In contrast, the numerical raw data and the summary chart was perceived as less helpful, as exemplified by the comment from P6 who thought
    \textit{``the numerical scores weren't really helpful; too often a low score wasn't justified in the open-ended feedback, and it was hard to understand what we could improve.''}
The numerical feedback was only \textit{``useful for finding the most obvious flaws and getting an overall feeling of how the people reacted''} (P27).
Nevertheless, students appreciated the summary chart, as they
\textit{``found it useful to have two different 
groups of people to evaluate our application [...] It forced us to think how we could make it more appealing to both groups''} (P43).


When evaluating the numerical feedback,
students primarily looked for differences between the two sources of feedback in the summary chart, not commonalities. ``Variety''
was mentioned several times as criteria to look for in the summary chart. P92, for instance, mentioned that \textit{``in the [Likert-scale] questions there was not enough variety to see which areas were good and which needed improvement.''} Students liked to contrast the feedback from the two sources, as it allowed them to identify the weaknesses and strengths in their designs.
For instance, P10 mentioned that
    \textit{``since both sources are not correlated, it was easy to identify the main design failures in the app and prioritize and solve them.''} 





\subsection{Perceived Fairness of the Feedback}%

Approximately two thirds of the students (62~students) found the peer feedback was more fair than crowdsourced feedback, compared to 34~students who thought the crowdsourced feedback was more fair (see Table~\ref{tab:codedquestions}).
Students primarily perceived the feedback's fairness along five dimensions:
    (a) \textit{agreeableness},
    (b) \textit{valence},
    (c) \textit{diligence},
    (d) \textit{usefulness}, and 
    (e) \textit{credibility}.
The five dimensions were mentioned by approximately the same number of students.

\subsubsection{Agreeableness}%
The students' perception of fairness was strongly determined by how agreeable they perceived the feedback to be.
Sensible and well justified feedback was easier to reconcile with their own view in this regard.
Constructive and well-thought out feedback was overall perceived to be more fair than feedback lacking these attributes.
Peer feedback was generally perceived to possess more of the above attributes than crowdsourced feedback.

Responses from students who
    preferred crowdsourced design feedback in terms of fairness
were often motivated by the lack of criticism which made the crowdsourced feedback ``more agreeable,'' such as the comment from P16 who found crowdsourced feedback was \textit{``much more positive and [workers] seemed to like our app. Fellow students were much more critical. I think our class tried to be too critical and find every little thing that was wrong.''}
Fairness, in the minority group who thought crowdsourced feedback was more fair, was simply a matter of the crowd finding \textit{``less problems with the application''} (P5).

Several students perceived feedback with a variety of different viewpoints as more fair. For these students, peer feedback contained a greater diversity of viewpoints and was thus judged to be more fair than crowdsourced feedback.

\subsubsection{Perceived valence}
As for the contribution of the affective tone (i.e., valence of the feedback) to the students' sense of fairness, students perceived the peer feedback as more critical, and much more harsh and negative in tone than the crowdsourced feedback. The criticism and negativity divided the students. While the majority of students appreciated critical feedback and perceived it as fair, a minority perceived the critical peer feedback as less fair than crowdsourced feedback.


Among the students who thought that critical peer feedback was fair, the crowdsourced feedback was often found to be overly positive and not critical enough to be useful. P1, for instance, found peer feedback more fair because
    \textit{``even the little feedback we got from [MTurk] was way too positive and approving and did not criticize the obvious problems the prototype had.''}
Similarly, P8 stated that he preferred peer feedback, because peers had
    \textit{``given more negative feedback compared to Mturk people and some of those comments helped lot to further improve and redesign our app.''}



Feedback from MTurk was generally perceived to be more approving and praiseful.
However, not all students considered positive feedback as being correlated with fairness.
Of the 16~students who mentioned MTurk feedback being positive, six found the crowdsourced feedback was fair, but 10 thought the crowdsourced feedback was unfair because of its overly positive praise.
%
%
As for the reasons of why peer feedback was more critical, P92 speculated that
    \textit{``working with our own designs during the course made us students more demanding towards each other's designs than the MTurk workers.''}

\subsubsection{Perceived diligence}
A high quantity of feedback positively contributed to the students' sense of fairness. The feedback from MTurk lacked in this regard and was perceived as less fair than the peer feedback due to the often short replies.
In the same vein, some students judged the MTurk feedback to be less fair because of its superficiality, lower clarity and lower specificity, as exemplified by the comments from
    P85 who thought that
    \textit{``peer feedback was more fair because the some crowdsourced feedback was not clear,''}
    and P8, who thought that
        \textit{``Mturk people gave more general comments which were useful too, while peers were more specific on errors and mistakes~-- like spellings.''} 
Internal consistency and absence of contradictions was mentioned by a few students as contributing positively to the fairness of the feedback.

\subsubsection{Perceived usefulness}
Feedback with a high number of suggestions
    and helpful hints
was perceived more fair than feedback lacking these features.
The effectiveness of the feedback and its perceived value also contributed to the students' sense of fairness.
Peer feedback overall was perceived more useful and fair than crowdsourced feedback.
One contributing factor to this sense of fairness was the context of the feedback's inception. The peers were familiar with the design task and had all undergone the same training. The peer feedback was thus perceived as more useful, as exemplified by the statement from P28:
\begin{quote}
    \textit{``the class seemed to know the context better, and gave more relevant feedback. Much of the MTurk feedback was useless.''}
\end{quote}
On the other hand, some students thought the crowdsourced feedback was more fair because of the crowd having an outsider's perspective on the application. P88, for instance, thought that
    \textit{``crowdsourcing was fairer because it felt like there was no obligation for the testers to be nice.''} 
The external reviewer having an impartial point of view also contributed to the students' perception of the credibility of the feedback, as detailed in the next section.

\subsubsection{Perceived credibility}%

The credibility of the feedback provider was an especially strong argument among the students who preferred the feedback from MTurk in terms of fairness.
Among these students, the feedback from external workers was perceived to be more objective than the feedback from classmates.
The \textit{``external and impartial point of view''} (P19) was appreciated by several students and contributed to the sense of credibility.
The workers were neither familiar with the course nor the students, and were thus perceived as being less biased, which positively contributed to the trust into the workers' feedback.


The majority of students, however, perceived peer feedback as more credible than crowdsourced feedback.
Among this majority, peer feedback was perceived as more serious and sincere, more realistic, and better scoped to the course.
Feedback from MTurk, on the other hand, was described as containing \textit{``a lot of joke answers''} (P15).
Some of the distrust into the crowdsourced feedback was, however, motivated by the students being unfamiliar with the external source of feedback. P47, for example, commented that he preferred
    \textit{``feedback from other students because I am not familiar with MTurk, so it is difficult to estimate the reliability of it.''} 



\subsection{Estimating the Monetary Value of Crowdsourced Feedback}%
\label{sec:sub:monetary-value}%

We asked each student to quantify the monetary value of the crowdsourced feedback for the group work (see Section~\ref{sec-sub:procedure}).
In this section, we provide an analysis of the responses to this item.

\subsubsection{Monetary value}%

We extracted the monetary value from each response (where possible and using the mid-point of a range in seven of the 106 responses).
Unrealistically high values (e.g., 5000 EUR) were removed using the upper 1.5~interquartile range (IQR) as a cut-off.
After the removal of outliers, the monetary value of the crowdsourced feedback ranged from 0~EUR to 25~EUR, with a mean value of 6.50~EUR
($Median=5$~EUR, $Q1=0.80$~EUR, $Q3=10$~EUR).


\subsubsection{Factors affecting the valuation}%

Twenty-nine of the 106 students provided written explanations for their choice of monetary value. In the remainder of this section, we provide an account of the 29 students' thoughts with specific attention to how students voiced and reasoned about their valuation of the feedback from crowd workers.
The overall sentiment of the feedback's value ranged from positive to extremely negative. 
Two main criteria for valuing the crowdsourced feedback emerged from the students' responses: perceived usefulness and professionalism (see Table~\ref{tab:valuecriteria}).
\rev{The valuation of the feedback fell in two camps:
    students who thought the feedback was completely worthless
        (we refer to them as \textit{``hardhats''} due to their strict, matter-of-factly and engineer-like approach to estimating the feedback's value), and
    students who attributed some value to the feedback
        (the \textit{``bungaloos''} due to their bona fide way of valuing the feedback driven by conflated expectations).}

\begin{table*}[htb]
\centering
\caption{Students' criteria for estimating the monetary value of crowdsourced feedback from MTurk.}
\label{tab:valuecriteria}
\begin{tabular}{ll}
\toprule
\td{\textbf{Criteria and sub-criteria}} & \td{\textbf{Description}}
\\
\midrule
    \td{\textbf{Perceived Usefulness}} & \td{The usefulness of the feedback and its appropriateness for completing the design task.} \\ 
        \td{\textbullet~Perceived Quality} & \td{The overall quality of the feedback.} \\
        \td{\textbullet~Perceived Relevance} & \td{The relevance of the formative feedback for the students' design task.} \\ 
        \td{\textbullet~Perceived Effort} & \td{The effort the worker put into completing the task. 
        } \\
        \td{\textbullet~Perceived Helpfulness} & \td{The amount of novel insights provided by the feedback to improve the design.} \\
        \midrule
        \td{\textbf{Professionalism}} & \td{The assumed expertise and skill set of the worker for conducting usability and user experience tests.} \\
\bottomrule
\end{tabular}
\end{table*}

\rev{Among the hardhats,} 
the bulk of the comments
mentioned not wanting to pay or not valuing the feedback at all. This was often justified with the low quality of the feedback.
P13, for example, would not pay anything \textit{``because the feedback was extremely poor.''}
Other students in this camp expressed their sense of value in more drastic ways, often motivated by the low usefulness and low relevancy of the crowdsourced feedback for the design task. P48, for instance, thought that
    \textit{``Mturk should pay us to read that rubbish, there were only two useful feedbacks and for them maybe few euros.''} 
Similarly, P28 complained that \textit{``only one comment was actually relevant''}, and 
\textit{``even if they paid you, this feedback would not be worth anything.''}
While the potential of the crowdsourced feedback was recognized, some students were disappointed, 
thinking that the feedback was worth
    \textit{``absolutely nothing, it was a disgrace and a waste of money for at least most of our feedback. I've seen some of our classmates in other groups getting fair feedback and specific ones but ours was just...blatantly saying extremely disappointing''} (P4).
P98 expressed regret that the crowdsourced feedback was worth \textit{``sadly nothing because it literally gave us nothing to work with.''}

A perceived low specificity of the crowdsourced feedback was a common theme in the responses from both camps, as exemplified by the comment of P89 (hardhat) who thought that some workers
\textit{``talked about the idea and not the design, and to top it all lot of the critics were not specific at all and so short we couldn't take anything from them (like `GOOD').''}
A suspicion of low effort 
affected the students' valuation of the feedback, as highlighted by P99 (bungaloo) who thought that some workers
    \textit{``didn't even spend 1 minute on filling the feedback form so those deserve 1€ max. We got a very detailed review also though which was worth 3-5€ imo.''} 


Only one of the 29~students explicitly mentioned considering the minimum wage of workers on MTurk. This student (P12, bungaloo) placed the value of the feedback \textit{``somewhere between whatever the minimum fee on MTurk was (90 cent or so I don't remember exactly) to few euros at max (maybe 5-6 if generous).''}

In the camp of bungaloos, the low amount of information in the feedback was recognized by several students, with also many complaints about the relevance and usefulness of the feedback.
P26, for instance, complained that \textit{``some of the feedback had basically no content in them''} and \textit{``it can't be more than maybe 30€ because if it's more than that, the whole thing is a rip off.''}
Similarly, P43 estimated the feedback from the nine workers to cost about 20 Euros, but was concerned whether the workers had
    \textit{``even evaluated the app or just thrown randomly points in the questionnaire.''} 

Many students from the bungaloo camp exhibited a na\"ivet\'e towards what an acceptable pay for the feedback would be, and how to value crowd work in general. For example, P88 noted that
    \textit{``the feedback helped streamline our app which I think is really valuable. A lot of the times small inconveniences can be a gamebreaker for products. Maybe 100€. I'm not quite shure [sic] how much feedback like that costs normally so its hard to put a concrete value to it.''} 
P102 admitted to \textit{``have no idea but I guess hundreds of euros, maybe closer to thousands.''}
P65~also noted that
    \textit{``I don't have proper knowledge about how much such online workers charge but i think 20--30 euro would be fine.''}

As is evident in some of the quotes above, students from the second camp often had a negative opinion of the feedback, but still considered paying a decent amount to the crowd workers.
P96, for example, complained that \textit{``all of the open ended questions were not useful,''} but thought the feedback was still worth 5€.
In the same vein, P32 mentioned the feedback was \textit{``really poor,''} but still considered paying \textit{``10~EUR per worker.''}

The primary contributor to this disaccord between perceived usefulness and monetary valuation was the student's mental image of the crowd worker as a trained professional with experience in usability and user experience (UX) testing.
The suspected professionalism strongly affected the students' valuation of the crowdsourced feedback:
    \textit{``I would only pay for one of the workers, maybe around 5-20 Euros depending on the expertise of the worker in this field''} (P9).
The poor quality of the feedback did not dissuade the students from thinking that the workers deserved to be paid, often with extremely generous amounts:
    \textit{``As they are professionals , if I consider the feedback on my application, the cost be between 100-200~Euros''} (P36).
Similarly, P68 valued the feedback with \textit{``150~[EUR] max,''} even though \textit{''some [open-ended] feedback was only numbers like `3'.''}%
%
%
%
%
%
%
\section{Discussion}%
\rev{The majority of students in our study perceived peer feedback as more useful, more detailed, more specific, more effective, more actionable, and of greater fairness than crowdsourced feedback from MTurk.
Further, the crowd workers spent less effort on writing their feedback, as evident in the high percentage of short answers. Peer feedback was more elaborate and contained less noise than the crowdsourced feedback.}

\rev{In line with prior work~\cite{p1024-nguyen.pdf},
the qualitative analysis of the final questionnaire revealed that students perceived the affective tone (valence) of anonymous feedback to be slightly more positive compared to feedback from peers. Peer feedback was perceived to be much harsher and critical in tone. This, however, did not negatively impact the students' sense of fairness. On the contrary, students appreciated critical feedback, as long as it was elaborate, specific, well justified, and useful.}
%
%
%
%
%
%
%
%
%
%
We found, however, a gap in how student's perceived and valued the crowdsourced feedback which may contribute to a false sense of achievement in students, as discussed in the following section.
%
%
%
%
\subsection{Perception Gap between the Monetary Value and Usefulness of the Feedback}%
\label{sec:sub:perceptiongap}%
\rev{Our data highlights that a student's perception of the monetary value of MTurk work may not be accurate.
But since students likely have no experience with crowdsourcing,
why is the students’ perception of feedback value relevant and interesting? Why does it matter if they know how much to pay crowd workers?}

\rev{From the students’ individual perspective in the design task, the monetary value of feedback may indeed be irrelevant.
From the perspectives of teaching, crowdsourcing research, and research funding, it is however exceedingly important to know that money on crowdsourcing is well spent as a monetary and pedagogical investment by the teaching organization. Knowing that students valued the feedback, and how they valued it, is therefore of particular importance to research and education institutions as well as teachers.}

Over the years, many researchers have argued for fair work conditions on crowdsourcing platforms (e.g., \cite{turkopticon.pdf,fairwork-hcomp2019.pdf}).
On the worker side, much of the discussion focuses on the imbalance in the dominant power structures on crowdsourcing platforms and on meeting a minimum wage for workers.
From the requester's perspective, the crowdsourcing literature primarily focuses on the design of effective quality control mechanisms.
Our work highlights a third side~-- that of the feedback receiver's qualitative experience of feedback.

The design space for crowdsourcing involving third parties has been only recently emerging.
Examples include PledgeWork~\cite{paper311.pdf}, a system for volunteers to donate their income from crowd work to a third party (a charity), revenue sharing (e.g., \cite{CrowdCOOP.pdf}),
and the review of subcontracting microwork by \citeauthor{subcontracting-crowdwork.pdf}~\cite{subcontracting-crowdwork.pdf}.
In the classroom with three parties (i.e., the teacher, crowd workers and students), the requester of feedback (the teacher) may not be the receiver and beneficiary of the feedback (in this case, the students).
This poses the question of how crowdsourced tasks for the collection of feedback should be priced.
As the feedback mainly contributes value to the feedback receiver, the feedback receiver may be the right party to determine the feedback's value and hence its price.
However, prior research on design feedback in the context of the classroom priced the crowdsourcing tasks 
without taking the feedback's value for students into consideration.


On the other hand, many students in our cohort were clearly unsure of how to value the crowdsourced feedback. Interestingly, these students still considered paying the workers a handsome amount of money for the feedback, even though the feedback was often of low usefulness. One reason for this gap in the perception and valuation of the feedback was that students conjured up an image of a professional and experienced online worker who is equipped with subject-specific expertise and an appropriate skill set. This conjured image, however, is in contrast to how MTurk is designed, as a marketplace for anonymous humans to complete short tasks irrespective of skills and expertise.
Further, local salary standards obviously skew people's understanding about the monetary value of labor on MTurk.

All things considered,
if teachers are to employ crowdsourced design feedback, teachers must educate their students about online work and its value.
We contend that in the age of the gig economy, crowdsourcing and especially its valuation 
should be a fixed part of the Computer Science curriculum.
\rev{Considering the strengths of employing crowdsourced feedback in the classroom, we concur with prior studies and argue that using crowdsourced feedback
as a complement to traditional feedback mechanisms is useful for students.}

\subsection{Complementing Peers with the Crowd}%

\rev{Our work confirms the findings of prior studies that crowdsourced feedback is a good supplement to peer feedback~\cite{p5580-wauck.pdf,crowdInnovationCourse-chi13.pdf}.}
While the students in our study generally favored peer design feedback, they also discovered and acknowledged clear value in the crowdsourced design feedback~-- value that is impossible to be obtained in the classroom setting alone. For instance, the diversity of people providing the feedback was mentioned.
In general, getting an outsider's perspective and feedback from diverse people from different cultures and backgrounds
 was valuable to the students. Crowdsourced design feedback was seen as a way to provide a reality check. Related to this, a few students stated that 
    crowd workers did not \textit{``hold back with their feedback''} (P67),
whereas many perceived peer feedback as being \textit{``somehow biased''} (e.g., P46, P100, and P105) even if they could not articulate this aspect in more detail.
Further, the students reported that contrasting the feedback from the two different sources helped them identify the weaknesses and strengths in their designs, which allowed the students to set priorities and take appropriate action to address the weaknesses in their designs. More specifically, the students could distinguish between the two different groups of application users, and think how to improve the application for both groups.
\rev{In the following section, we reflect on our own perspective as teachers of the HCI course.}

\subsection{Reflections from the Teachers' Perspective}%

\rev{The complementary value of crowdsourced feedback becomes a question of trade-offs between the added value to the learning versus the added burden to teachers and the monetary cost of the feedback.
In our experience\footnote{Several of the article's authors were involved in organizing this study's course.} there is certainly something of value in students seeing their designs being rated by other than familiar faces in the class: it is exhilarating and introduces an element of excitement to teaching.
From the teachers' standpoint, utilizing MTurk as a feedback source was a refreshing experiment and provided us a chance to expose the students to feedback from people who are not their ``friendly peers.'' 
Further, and while we have no data to back this up, we hypothesize that the knowledge of one's design ending up online and being inspected by others than peers motivates the students to put in more effort.}


\rev{Ultimately, teachers are the ones who choose to use crowdsourced feedback or not. In our case, for example, this will certainly not be the last time we employ crowd workers for additional feedback.
Yet, several questions to consider remain. For instance, what is the correct way to determine the value of crowdsourced feedback? In the case of crowdsourcing platforms, the feedback is simply work that has a certain monetary value, dictated by the platforms' rules as well as the workers' and requesters' perceptions. 
The classroom as a context distorts this simple way of looking at value, however.
In many institutions teaching does not have any extra budget available, and the question of value is not just about money. 
It is also a question of time and effort required to set it all up: collecting and distributing the feedback. We did this manually, and it certainly was not an insignificant amount of work.
Not all teachers are equipped with the skills and knowledge required to crowdsource feedback from the crowds, and learning such skills could be difficult. There is a dearth of tools that would ease the setup for non-technically savvy teachers, and given how different each case most likely is, it is challenging to even envision sufficiently generic and easy-to-use tools for crowdsourcing feedback that would work across all types of classes.}

\rev{Further, with teaching one has to always consider the quality of feedback.
Instructors supposedly are experts in the subjects they teach, and the students can thus trust the feedback and its quality. 
With crowds, there is always an uncontrolled element of uncertainty involved. 
Should we trust that the crowd feedback, collected and distributed as-is, is of sufficient quality, or should the instructors vet and double-check everything? Would this, then, lead to loss of authenticity in the crowd feedback, even if authenticity is one of the perks of it? 
We believe that a fully transparent approach might work best here; Collecting the feedback from crowds but at the same time explaining the students exactly what they are in for and from what type of workforce.}




\rev{We envision students to take a more active role in collaboratively working with crowds in the classroom, and we have
already taken this approach in our own teaching.
Our work may inspire the design of pedagogical tools and interventions (e.g., a set of rubrics), using the criteria that we outline in the paper as a framework.
For example, we now encourage students to define their own set of rubrics and guiding questions for collecting feedback as part of their learning experience.
To this end, we give recommendations for using crowdsourced feedback in the classroom in the next section.}

\subsection{Recommendations for Crowdsourcing Design Feedback in the Classroom}%
\rev{In this section, we consolidate recommendations that are common knowledge in the crowdsourcing community and available in the crowdsourcing literature (e.g.,~\cite{crowdInnovationCourse-chi13.pdf}), but may be novel for teachers in the HCI community who are less familiar with crowdsourcing and wish to use crowdsourcing as a source of feedback in the classroom.}

\subsubsection{Set the right expectations in students}%
    We contend that teachers should educate their class about crowdsourcing (and crowd work in general), and especially about the working conditions on crowdsourcing platforms, and how such work is commonly priced.
    To effectively evaluate the crowdsourced feedback, students must be aware of the strengths and weaknesses of the crowdsourcing model.
    In the context of a course that provides crowdsourced feedback to students, educators should collect demographic data about the crowd workers' educational and professional background. This demographic data may help students to adjust their conflated expectations and arrive at a more realistic valuation of the crowdsourced feedback.

\subsubsection{Set the right expectations in crowd workers}
    According to feedback intervention theory, if feedback signals that an effort falls short of an expected standard, learners become motivated to increase their efforts to attain the standard~\cite{e81ca813ed757e1e76c0023865c7dbdc7308.pdf}.
    To implement this mechanism successfully, the standard and expectations must be clearly explained to the feedback provider.    
    The crowd workers should be made aware that their feedback is given to students as formative, not summative feedback.
    Feedback requesters should define clear success criteria for the crowd workers and provide a definition of what defines good feedback.
    Structuring feedback with rubrics, scaffolds, guiding questions, and other structured work flows may help increase the quality of the feedback in this regard \cite{paper138.pdf,3357236.3395480.pdf,Posts_paper_3.pdf}. 
    In particular, requesters of crowdsourced feedback should pay attention to how the criteria that affect the students' perception of quality
    and fairness
    shape the students' experience of feedback. Critical and harsh feedback, for instance, was not necessarily perceived as unfair by the students.


\subsubsection{Apply crowdsourcing best practices of quality control}%
    To effectively complement each other, both sources of feedback must contribute some valuable insights.
    To this end, typical qualification criteria as employed in our study (i.e., 95\% past acceptance rate and 100 completed HITs) did not prove to be enough to motivate the MTurk workers to provide good feedback.
    To this end, the quality of the crowdsourced feedback could be elevated by filtering responses (post data collection). This is indeed a common practice with crowdsourced data collection.
    In our study, we found that the open-ended feedback provided by the crowd workers markedly improved when short responses were removed. The relevance of the feedback and its usefulness can be expected to improve if these responses are discarded.
    



\subsubsection{Use an appropriate feedback modality and consider feedback aggregation}%
    Another consideration is how feedback should be collected and provided to students.
    %
    %
    As is evident in a number of crowd feedback systems (e.g., \cite{p1433-xu.pdf,CrowdUI,3313831.3376380.pdf}),
    visualization and aggregation of feedback supports the feedback receiver in making sense of the feedback.
    Research on mechanisms for aggregating crowdsourced design feedback 
    is only recently emerging (e.g., \cite{3313831.3376380.pdf}).
    Our study found that students in particular valued diversity in the responses and appreciated the direct contrast between the feedback from the two sources.%
%
%
%
%
\subsection{Limitations}%
We acknowledge limitations in our study.
First, the feedback receiver's pedagogic literacy affects how feedback is evaluated~\cite{595a3a43a246214055c494fe3f0e899f7bb4.pdf}.
Our findings may therefore be specific to the class and we do not claim that the findings generalize to other cohorts of students or other study subjects.

Second, the subjective experience of feedback is influenced by a number of factors.
For instance, the order in which feedback is received may affect the perception of feedback~\cite{p137-wu.pdf,p1024-nguyen.pdf}.
    \rev{\citeauthor{p1024-nguyen.pdf} found that framing feedback for a writing task with positive affective language had a positive effect on work quality~\cite{p1024-nguyen.pdf}.
    In \citeauthor{p137-wu.pdf}'s study, participants were more motivated and perceived the feedback as most favorable when negative feedback was given after positive feedback~\cite{p137-wu.pdf}.}
In our study, we did not control the order of feedback.
\rev{Students explored the contents of the feedback package on their own terms.}
We argue that our study setup is not unrealistic, as it aligns well with the microtask crowdsourcing model found on MTurk and reflects how feedback could practically be provided in the classroom.


Third, a limitation of the investigation into the monetary valuation of crowdsourced design feedback is that while the questionnaire item 
specifically asked
the students to estimate the value of the crowdsourced feedback as a whole, some students may still have estimated the value per worker. However, the students who elaborated on their answer typically mentioned whether their estimate was per worker or for the whole of the crowdsourced feedback, as discussed in Section~\ref{sec:sub:monetary-value}.


\rev{Fourth, knowledge of the feedback source may have influenced the students’ interpretation of the feedback. The choice to disclose the source was, however, imperative for teaching students the value of the two feedback sources. We believe revealing the source at the end of the study would have created tension among the students.
Further, peer feedback was not assigned a value in our study.
Peer feedback in our context (i.e., teaching an HCI class) is a mandatory part of the course and a traditional element of teaching and learning. 
In our institution’s ethical stance, peer feedback can and should not be assigned a monetary value, and taxation law prohibits us from handing out money to students.
We did, however, not see signs of insincerity in the students’ answers (after removing the few unrealistic responses -- see Section 5.3.1).
We acknowledge that alternative study designs could have been explored.}

Last, we did not control the students' familiarity with crowdsourcing.
\rev{Students received a general introduction to crowdsourcing in the lectures (but without going into detail about the work reality on crowdsourcing platforms).}
A few students had prior knowledge of the work conditions on crowdsourcing platforms and were thus able to accurately estimate the price of crowdsourced work. The majority of the students, however, was new to crowdsourcing, as evident in their responses to our survey. 


\section{Conclusion}%


In this work, we provided a detailed  empirical 
study  of how students perceive and value crowdsourced feedback, as well as how crowdsourced feedback compares to peer feedback in the classroom.
We found that students preferred peer feedback as it was perceived as more
useful, fair, effective, and actionable.
Additionally, our investigation of the students' monetary valuation of the crowdsourced feedback revealed quality, relevancy, effort, and helpfulness as important factors that shape the value of the feedback for students.
We found clear evidence that some students were na\"ive toward
    the work conditions on crowdsourcing platforms and how such work is priced.
The monetary valuation of the crowdsourced feedback by these students was strongly shaped by a mental image of the worker as a trained professional.
Ultimately, we believe that crowdsourced design feedback in HCI teaching provides a great way to complement peer feedback, as long as student expectations are calibrated adequately.

\section*{Acknowledgments}%
\rev{We thank the students of the HCI 
course and the workers from Amazon Mechanical Turk who contributed their feedback in this study. This research is partially enabled by the GenZ strategic profiling project at the University of Oulu, supported by the Academy of Finland (project number 318930).}


\bibliographystyle{ACM-Reference-Format}
\bibliography{main}

\end{document}